# A Lattice Study of Semi-leptonic Decays of $D$-Mesons

$UKQCD$ $Collaboration$ - presented by J. Nieves[a][*]

[a]Physics Department, The University, Southampton SO17 1BJ, UK

We present results of a lattice computation of the matrix elements of the vector and axial-vector currents which are relevant for the semi-leptonic decays $D \to K$ and $D \to K^*$. The computations are performed in the quenched approximation to lattice QCD on a $24^3 \times 48$ lattice at $\beta = 6.2$ using an $O(a)$-improved fermionic action.

## 1. Introduction and Simulation Details

There is increasing evidence that quantitative calculations of weak decay amplitudes can be obtained by lattice QCD simulations. $D$ decays provide a good test of the method, since the relevant CKM matrix elements ($V_{cs}$ and $V_{cd}$) are well constrained in the Standard Model. In addition comparisons between $D$ and $B$ decays reveal the size of non-leading terms in the Heavy Quark Effective Theory (HQET). The main advantage of the lattice technique is that it is based on first principles only and it does not contain free parameters besides the quark masses and the value of the lattice spacing, both of which are fixed by hadron spectroscopy. Moreover, statistical and systematic errors in lattice simulations can be systematically reduced with increasing computer resources. In this study we present the results of a lattice calculation of the matrix elements $\langle K^*|J^\mu|D\rangle$ and $\langle K|J^\mu|D\rangle$ ($J^\mu = \bar{s}\gamma^\mu(1-\gamma_5)c$ is the relevant quark weak current) which contain the non-perturbative QCD corrections to the decays $D \to K l^+ \nu_l$ and $D \to \bar{K}^* l^+ \nu_l$. We have used an improved quark action, proposed by Sheikholeslami and Wohlert [1], which together with an appropriate rotation of the quark fields, leads to matrix elements free of discretisation errors of order $a\alpha_s^n \log^n(a)$. When using this action the leading discretisation errors are of the order $O(\alpha_s a)$ and $O(a^2)$ [2]. This "improvement" is particularly important here since we are studying the propagation of quarks whose bare masses are around one third the inverse lattice spacing. We have analysed 60 gauge field configurations at $\beta = 6.2$ on a lattice of size $24^3 \times 48$. Propagators were computed at three "light" values of the hoping parameter, $\kappa = 0.14144, 0.14226, 0.14262$, and at $\kappa_c = 0.129$ which corresponds to the mass of the charm quark. In order to determine the matrix elements mentioned above, we compute three-point correlation functions in which the $D$ meson is placed at $t = 24$, the midpoint of the lattice, and symmetrize the correlators about that point using Euclidean time reversal. We give momenta $|\vec{p}_D| = 0, \frac{\pi}{12}a^{-1}$ to the $D$-meson and momentum $\vec{q}$ is inserted at the current. We use the local, improved vector current. Results are extrapolated to the physical values of the strange ($\kappa_s = 0.1419(1)$) and light quark masses ($\kappa_{crit} = 0.14315(2)$). The "light" quark propagators were computed using local sources and sinks; the "charm" propagator was computed using gauge-invariant "smeared" source and both local and smeared sinks. Full details of the calculations are contained in ref. [3].

## 2. Renormalisation Constants $Z_V$ and $Z_A$

In this section we discuss the difficulties in determining the form factors of semi-leptonic $D \to K, K^*$ decays due to the presence of discretisation errors. In general the matrix elements of an "improved"-lattice current, $V_\mu^L$, are related to the corresponding continuum matrix elements by:

$$\langle f|V_\mu|i\rangle^{continuum} = Z_V(\alpha_s)\langle f|V_\mu^L|i\rangle +$$

[*]Talk presented at Lattice'94, Bielefeld, Germany, 27 Sep-1 Oct 1994. This research was supported by the UK Science and Engineering Research Council under grants GR/G 32779 and GR/J 21347, by the European Union under contracts CHRX-CT92 − 0051 and CHBICT920066 (postdoctoral fellowship), by the University of Edinburgh and by Meiko Limited.

$$O(\alpha_s a) + O(a^2) + \ldots \quad (1)$$

where the renormalisation constant $Z_V$ is independent of the external states $i$ and $f$, and is computable in perturbation theory. For light quark masses the discretisation errors are small but for the charmed quark this is no longer the case. In previous simulations, using Wilson fermions (for which there exist $O(a)$ discretisation errors not present in this simulation), these effects were modelled by using effective (mass-dependent) renormalisation constants $Z_V^{eff}$ and $Z_A^{eff}$. However we wish to stress that the discretisation errors are in general different for matrix elements of currents with different Lorentz indices, between different states, and they could have a different $q^2$ dependence than the form factors themselves. Formally there is no reason to believe that the discretisation errors can be absorbed into universal effective renormalisation constants. In ref. [3] the size of the lattice artifacts in the present study is discussed, and the conclusion is that although these errors are substantially reduced by the use of the improved action, nevertheless even in this case we believe that they lead to uncertainties of the order of 10% in the form factors. In spite of the above discussion we have also assumed that they can be modelled by effective renormalisation constants $Z_V^{eff}$ and $Z_A^{eff}$ (at least part of the errors can be so absorbed).

## 3. Results

In Figs. 1 and 2 we show the form factors $f^+$ and $A_1$ as a function of $q^2$ for the semi-leptonic decays $D \to K, K^*$. In both cases the solid line corresponds to the comparison of the pairs ($q^2$, form factor) with the $q^2$-dependence of the form factors determined from a two-parameter (form factor at $q^2 = 0$ and pole mass) pole dominance fit to our data. Crosses correspond to the form factors at $q^2 = 0$ (up to factors $Z_V^{eff}$ and $Z_A^{eff}$ respectively) determined in this way. In the case of $f^+$ we also compare (dashed line) the $q^2$-dependence of our data with that determined from a one-parameter pole dominance fit, where we fix the pole mass to that of the vector-meson $D_s^*$, which is the resonance in this helicity channel and which is also in an excellent agreement with the experimental determination of the $q^2$ dependence of $f^+(q^2)$ by the CLEO Collaboration ([4]).

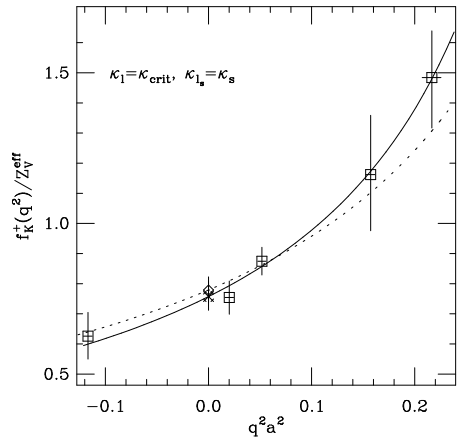

Figure 1. $f_K^+(q^2)$ as a function of $q^2 a^2$.

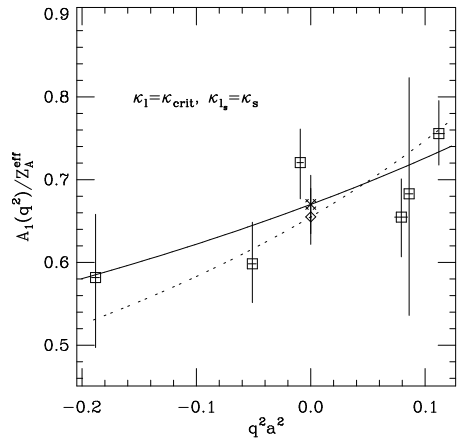

Figure 2. $A_1(q^2)$ as a function of $q^2 a^2$.

For the axial form factor, $A_1$, the dashed line also corresponds to a one-parameter pole dominance fit, where the pole mass is fixed to 2.5 GeV corresponding to the mass of the axial-vector resonance ($D_{s1}$) predicted in this helicity channel by quark models. In both figures diamonds correspond to the form factor at $q^2 = 0$ (up to factors $Z_V^{eff}$ and $Z_A^{eff}$ respectively) determined by using one-parameter pole dominance fits. The fits using the constrained pole masses also lead to acceptable $\chi^2/dof$. As can be seen from these figures both methods of extracting the form factors at $q^2 = 0$ agree remarkably well, which gives us confidence in our procedure. It can be also seen from these figures that the $q^2$ dependence of both form factors is reasonably well described by the pole dominance model. Qualitatively similar conclusions are obtained for the rest of form factors: $f^0, A_0, V$ and $A_2$ (for details see ref. [3] where results for the form factors relevant to the decays $D \to \pi, \rho$ are also given). In Table 1 we show our results for the form-factors at $q^2 = 0$ for the semi-leptonic decays $D \to K, K^*$. In our quoted errors we have included the uncertainty in the renormalisation constants, discussed in the previous section, in order to take into account some of the residual discretisation errors. In this table we also compare our predictions with the most recent experimental world average and with previous lattice results. Discretisation errors are, in principle, larger for Wilson than for improved actions and part of the discrepancies between different lattice results in Table 1 are due to different values used in the literature for the effective renormalisation constants $Z_V^{eff}$ and $Z_A^{eff}$. We have also found the $q^2$ dependence of the form factors in a wide region around $q^2 = 0$. Thus we can estimate the phase space integrals and obtain the total decay rates (Table 2). In both cases (Tables 1 and 2) our results are in good agreement with experimental data. The analysis of the $B \to \pi, \rho$ matrix elements is in progress.

Table 1
Form factors at $q^2 = 0$ for the semi-leptonic decays $D \to K$ and $D \to K^*$.

|  |  | $f_K^+(0)$ | $A_1(0)$ | $V(0)$ | $A_2(0)$ |
|---|---|---|---|---|---|
| Exp. | World Ave. [4] | $0.77 \pm 0.04$ | $0.61 \pm 0.05$ | $1.16 \pm 0.16$ | $0.45 \pm 0.09$ |
|  | World Ave. [5] | $0.70 \pm 0.03$ |  |  |  |
| $\beta = 6.2$, SW-$O(a)$ | This work | $0.67^{+7}_{-8}$ | $0.70^{+7}_{-10}$ | $1.01^{+30}_{-13}$ | $0.66^{+10}_{-15}$ |
| $\beta = 6.4$, Wilson | ELC [6] | $0.60^{+15}_{-15}{}^{+7}_{-7}$ | $0.64 \pm 0.16$ | $0.86 \pm 0.24$ | $0.40^{+28}_{-28}{}^{+4}_{-4}$ |
| $\beta = 6.0$, Wilson | LMMS [7] | $0.63 \pm 0.08$ | $0.53 \pm 0.03$ | $0.86 \pm 0.10$ | $0.19 \pm 0.21$ |
| $\beta = 6.0, 5.7$, Wilson | BKS [8] | $0.90^{+8}_{-8}{}^{+21}_{-21}$ | $0.83^{+14}_{-14}{}^{+28}_{-28}$ | $1.43^{+45}_{-45}{}^{+48}_{-49}$ | $0.59^{+14}_{-14}{}^{+24}_{-23}$ |
| $\beta = 6.0$, Wilson | BG [9] | $0.73 \pm 0.05$ | $0.66 \pm 0.03$ | $1.24 \pm 0.08$ | $0.42 \pm 0.17$ |
| $\beta = 6.3$, Wilson | WU [10] | $0.76 \pm 0.15$ | $0.59 \pm 0.08$ | $1.05 \pm 0.33$ | $0.56 \pm 0.40$ |
| $\beta = 6.0$, SW-$O(a)$ | APE [11] | $0.78 \pm 0.08$ | $0.67 \pm 0.11$ | $1.08 \pm 0.22$ | $0.49 \pm 0.34$ |

Table 2
Semi-leptonic widths for $D \to K, K^*$ decays.

|  | $\Gamma(K)$ | $\Gamma(K^*)$ | $(\frac{\Gamma_L}{\Gamma_T})_{K^*}$ |
|---|---|---|---|
| Exp. [4] | $9.0 \pm 0.5$ | $5.1 \pm 0.5$ | $1.2 \pm 0.1$ |
| Exp. [5] | $7.1 \pm 0.6$ | $4.5 \pm 0.5$ |  |
| This work | $7.0 \pm 2.0$ | $6.0^{+0.8}_{-1.6}$ | $1.1 \pm 0.2$ |

I thank my colleagues from the UKQCD Collaboration and C. Allton, T. Bhattacharya, S. Güsken, V. Lubicz and G. Martinelli for fruitful discussions.

## REFERENCES

1. B. Sheikholeslami and R. Wohlert, Nucl. Phys. B259 (1985) 572.
2. G. Heatlie et al., Nucl. Phys. B352 (1991) 266.
3. UKQCD Collaboration, K.C. Bowler et al., Edinburgh Preprint: 94/546, and hep-lat/9410012.
4. M.S. Witherell, talk given at the "International Symposium on Lepton and Photon Interactions at High Energies", (August 1993).
5. G. Bellini, talk given at "Les Rencontres de Physique de la Vallee d'Aoste", (March 1994).
6. A. Abada, et al. Nucl. Phys. B416 (1994) 675.
7. V. Lubicz, et al. Phys. Lett. B274 (1992) 415.
8. C. Bernard, A. El-Khadra and A. Soni, Phys. Rev. D43 (1991) 2140; D45 (1992) 869.
9. T. Bhattacharya, in these proceedings.
10. S. Güsken, in these proceedings.
11. APE Collaboration, C.R. Allton et. al., ROME prep. 94/1050 and hep-lat/9411011.